%% file: RXMHD_wavesV2.tex
\documentclass[a4paper]{cas-dc}

\input{preamble}

\begin{document}
\let\WriteBookmarks\relax
\def\floatpagepagefraction{1}
\def\textpagefraction{.001}

\journal{Physics Letters A} 

\shorttitle{Waves in Relativistic Extended MHD}
\shortauthors{Y. Kawazura}

\title [mode = title]{Yet Another Modification of Relativistic Magnetohydrodynamic Waves: Electron Thermal Inertia}

\author[1,2,3]{Yohei Kawazura}[type=editor, auid=000, bioid=1, orcid=0000-0002-8787-5170]
\cormark[1]
\cortext[cor1]{Corresponding author}
\ead{kawazura@tohoku.ac.jp}


\address[1]{Frontier Research Institute for Interdisciplinary Sciences, Tohoku University, 6-3 Aoba, Aramaki, Sendai 980-8578, Japan}
\address[2]{Department of Geophysics, Graduate School of Science, Tohoku University, 6-3 Aoba, Aramaki, Aoba-ku, Sendai 980-8578 Japan}
\address[3]{Astrophysical Big Bang Laboratory, RIKEN, 2-1 Hirosawa, Wako, Saitama 351-0198, Japan}

\begin{abstract}
This study investigates the properties of waves in relativistic extended magnetohydrodynamics (RXMHD), which includes Hall and electron thermal inertia effects. 
We focus on the case when the electron temperature is ultrarelativistic, and thus, the electron thermal inertia becomes finite at near the proton inertial scale.
We derive the linear dispersion relation of RXMHD and find that the Hall and electron thermal inertia effects couple with the displacement current, giving rise to three superluminous waves in addition to the slow, fast, and Alfv\'en waves.
We also show that the phase- and group-velocity surfaces of fast and Alfv\'en waves are distorted by the Hall and electron thermal inertia effects.
There is a range of scales where the group velocity of fast wave is smaller than that of the Alfv\'en and slow waves.
These findings are applicable to a region near the funnel base of low-luminosity accretion flows where electrons can be ultrarelativistic.  
\end{abstract}


\maketitle

\section{Introduction} \label{s:introduction}
Magnetohydrodynamics (MHD) is undeniably the most widely used model for understanding large-scale dynamics in astrophysics.
While it is a minimal extension of hydrodynamics to include electromagnetic fields, MHD can explain a variety of plasma phenomena.
However, MHD is valid only in a certain parameter range in which any microscopic effects are negligible.
To interpret modern astronomical observations, one must carefully consider the microscopic effects.
For example, the black hole shadow image at M87 captured by the Event Horizon Telescope~\cite{EHT2019a} can be theoretically explained in multiple possible scenarios depending on how the microscopic effects are treated~\cite{EHT2019e}.
As such, it is important to extend the MHD theory so that it includes microscopic effects self-consistently.

Such an attempt has been achieved in a non-relativistic regime by developing the extended MHD (XMHD)~\cite{Kimura2014,KeramidasCharidakos2014,Abdelhamid2015,Lingam2016}, which has two essential microscopic effects:
the Hall effect appearing at the ion skin depth $d_\rmi$ and the electron-rest-mass-inertia effect appearing at even smaller electron skin depth $d_\rme$.
For the last few years, XMHD has been used to study a variety of processes, such as magnetic reconnection~\cite{Grasso2017}, nonlinear waves~\cite{Abdelhamid2016a,Abdelhamid2017a}, and turbulence~\cite{Abdelhamid2016b,Miloshevich2017,Miloshevich2018}.

Meanwhile, relativistic effects are critical to understand high-energy astrophysical phenomena.
The relativistic XMHD (RXMHD)\footnote{
One may confuse RXMHD with the model with the same name developed by Chandra et al.~\cite{Chandra2015}. 
The latter model is essentially a relativistic version of Braginskii's equations~\cite{Braginskii1965} which take into account of weakly collisional effects, such as anisotropic pressure and heat flux but does not include microscopic effects.
On the other hand, RXMHD used in this study has microscopic effects, but no collisionless effects are included.
}
was formulated using an empirical approach~\cite{Koide2009,Koide2010} and a variational approach~\cite{Kawazura2017a}, and  it has been actively used for the last few years~\cite[e.g.,][]{Asenjo2015,Yang2016,Yang2018,Liu2018,Comisso2018,Liu2019,Asenjo2019,Comisso2020,Koide2020}. 
In RXMHD, the electron thermal inertia (ETI) effect emerges when the electron thermal energy exceeds the rest mass energy.
Remarkably, the scale at which ETI switches on increases as the electron temperature increases, and it can be significantly larger than $d_\rme$ where the electron rest mass inertia becomes finite. 
One of the most important consequences of the increased electron inertial scale is collisionless reconnection happening at a larger scale than $d_\rme$~\cite{Comisso2014,Comisso2018}.

Surprisingly, despite the obvious success of RXMHD, the most basic characteristic of RXMHD, namely the properties of linear wave propagation, has yet to be studied.
Knowing the linear wave properties has wide implications, including the development of direct numerical simulation codes~(see, for example, \cite{Komissarov1999,Balsara2001,DelZanna2003,Mignone2006,Giacomazzo2007} for relativistic MHD codes and \cite{Amano2015} for a non-relativistic XMHD code).
It is known that the properties of linear waves in non-relativistic and relativistic ideal MHD are quite similar~\cite{Keppens2008} except that the phase speeds of the three MHD waves, Alfv\'en, slow- and fast-magnetosonic waves, are bounded by the speed of light for relativistic MHD.
On the other hand, the non-relativistic Hall effect changes the wave properties dramatically~\cite{Hameiri2005}
For example, the phase and group diagrams (also known as Friedrichs diagrams) are distorted by the Hall effect, and the fast wave becomes strongly anisotropic, i.e., it primarily propagates in the direction parallel to the background magnetic field.  
In our previous work~\cite{Kawazura2017b}, we explored the wave properties in relativistic Hall MHD (RHMHD), in which the electron temperature is modestly- or non-relativistic, and thus ETI is negligible in ion inertial scales.
We demonstrated that relativistic Hall effect and non-relativistic Hall effect change the wave properties differently. 
This is a remarkable difference considering that waves are almost the same in non-relativistic and relativistic ideal MHD.
In RHMHD, the fast wave is more isotropic than that in non-relativistic HMHD, and the group velocity surface of the fast wave overlaps with that of Alfv\'en wave. 
Furthermore, the scale at which the Hall effect switches on is different from $\delta_\rmi$ when the relativistic effect is present.
This scale shrinks as the magnetic energy increases.
In other words, RHMHD becomes relativistic ideal MHD when the magnetic field is strong.

In this study, we additionally include ETI at ion inertial scales by considering electrons with ultrarelativistic temperature and investigate how ETI changes the wave properties of RHMHD.
It is suggested by the global simulations of black hole accretion flows that such ultrarelativistic electrons may exist in the funnel base region of low-luminosity accretion flows~\cite{Ressler2017}.\footnote{
Note that the electron temperature calculated by the global simulations strongly depends on the prescription of energy partition between ions and electrons~\cite{Moscibrodzka2016}, which is determined physically by the collisionless mechanisms. 
While the electrons become ultrarelativistic when the turbulent heating model~\cite{Howes2010,Kawazura2019,Schekochihin2019} is employed, the electron temperature becomes much lower with the reconnection heating model~\cite{Rowan2017,Rowan2019}.
Even with the turbulent heating model, the electron temperature can be lower if there are compressive fluctuations~\cite{Kawazura2020,Zhdankin2021}.
}
Hence, the wave properties revealed here will be useful to understand relativistic jets in active galactic nuclei.

\section{Linear Dispersion Relation of RXMHD} \label{s:model}
We consider a proton-electron plasma in a flat spacetime defined by the Minkowski metric, $\mr{diag}(1, -1, -1, -1)$.
The set of RXMHD equations~\cite{Koide2009, Koide2010, Kawazura2017a} are the continuity equation
\begin{subequations}
\begin{equation}
  \p_\alpha\lf( nu^\alpha \ri) = 0, 
  \label{e:XMHD continuity} 
\end{equation}
the momentum equation
\begin{multline}
  \p_\alpha\bigg[ nhu^\alpha u^\beta + \f{1}{e}(\mu h_\rmi - h_\rme)(u^\alpha J^\beta + J^\alpha u^\beta) \\
  + \f{1}{ne^2}(\mu^2 h_\rmi + h_\rme)J^\alpha J^\beta \bigg] = \p^\beta p + J^\alpha {F^\beta}_\alpha, 
  \label{e:XMHD e.o.m.} 
\end{multline}
the generalized Ohm's law
\begin{multline}
  \p_\alpha\bigg[ n(\mu h_\rmi - h_\rme)u^\beta u^\alpha + \f{1}{e}(\mu^2 h_\rmi + h_\rme)(u^\beta J^\alpha + J^\beta u^\alpha) \\
  + \f{1}{ne^2}(\mu^3 h_\rmi - h_\rme)J^\beta J^\alpha \bigg] = \mu\p^\beta p_\rmi - \p^\beta p_\rme  \\
  + enu^\alpha {F^\beta}_\alpha - J^\alpha {F^\beta}_\alpha,
  \label{e:XMHD Ohm's law} 
\end{multline}
and Maxwell's equation:
\begin{equation}
  \p_\alpha F^{\alpha\beta} = 4\pi J^\beta, \quad \p^\alpha(\epsilon_{\alpha\beta\gamma\delta}F^{\gamma\delta}) = 0,
  \label{e:Maxwell} 
\end{equation}
\end{subequations}
where $e, n, h_s, p_s,$ and $\epsilon_{\alpha\beta\gamma\delta}$ are the elementary charge, rest frame number density, thermal enthalpy, thermal pressure, and Levi-Civita symbol, respectively.
The subscript $s = (\rmi,\, \rme)$ denotes the species label with i for protons and e for electrons.
We also use the total enthalpy $h = h_\rmi + h_\rme$ and total pressure $p = p_\rmi + p_\rme$.
The four vectors and tensor represent the four velocity $u^\alpha = (\gamma,\, \gamma \bm{v}/c)$, the Faraday tensor $F^{\alpha\beta}$ that gives electric field $E_i = F_{0i}$ and magnetic field $B_i = -(1/2)\epsilon_{ijk}F^{jk}$, and four current $J^\alpha = \p_\beta F^{\beta\alpha} = (\rho_q,\, \bm{J}/c)$, where $\gamma = 1/\sqrt{1 - v^2/c^2}$ is the Lorentz factor of bulk flow, and $\rho_q$ is the charge density. 

One finds quite a few additional terms in \eqref{e:XMHD e.o.m.} and \eqref{e:XMHD Ohm's law} compared to the relativistic ideal MHD.
The final term in the right hand side of \eqref{e:XMHD Ohm's law} is a Hall term, while the terms multiplied by $h_\rme$ and $\mu$ in \eqref{e:XMHD e.o.m.} and \eqref{e:XMHD Ohm's law} are originated from the electron thermal and rest mass inertiae.
In our previous paper~\cite{Kawazura2017b}, we neglected the electron inertia effects and focused on the Hall effect by assuming that the electron temperature is modestly- or non-relativistic, i.e., $T_\rme/m_\rme c^2 \lesssim 1$, and wavelength is in proton inertial scales, i.e., $kd_\rmi \sim 1$, where $d_\rmi = (m_\rmi c^2/4\pi n_0 e^2)^{1/2}$ is the proton skin depth.
In this study, we consider electrons with ultrarelativistic temperature while the wavelength is slightly shorter than $d_\rmi$ but much larger than the electron inertial length $d_\rme = \mu^{1/2}d_\rmi$.
More specifically, we impose the following ordering
\begin{multline}
  T_\rmi/m_\rmi c^2 \lesssim 1, \; T_\rme/m_\rme c^2 \sim \mu^{-1/2}, \; \p^\alpha \sim \mu^{-1/4}d_\rmi^{-1},\\ u^\alpha \sim 1,\; F^{\alpha\beta} \gtrsim (nm_\rmi c^2)^{1/2}.
  \label{}
\end{multline}
The first, second, and fourth conditions are numerically supported to be valid around the funnel base region~(see, for example, Fig.~4 in Ref.~\cite{Sadowski2017} for the first condition, Fig.~1 in Ref.~\cite{Ressler2017} for the second condition, and Fig.~6 in Ref.~\cite{White2020} for the fourth condition). 
These conditions lead to $\beta = 8\pi n (T_\rmi + T_\rme)/B_0^2 \lesssim 1$ and $\sigma = B_0^2/4\pi n_0 h_{\rmi0} \gtrsim 1$, both of which means that the plasma is magnetically dominated, and they are valid around the funnel base region.

We then linearize \eqref{e:XMHD e.o.m.} and \eqref{e:XMHD Ohm's law} by splitting the fields as $f = f_0 + \tilde{f}$ where the subscript 0 denotes the spatio-temporary constant background field and tilde denotes the perturbation.
We also assume that there is no background flow, i.e., $\bm{v}_0 = 0$.
The background part of Maxwell’s equations \eqref{e:Maxwell} gives $\rho_{q0} = 0$, $\bm{E}_0 = 0$, and $\bm{J}_0 = 0$. 
Dropping the terms with the order higher than $O(\mu^0)$, the spatial parts of \eqref{e:XMHD e.o.m.} and \eqref{e:XMHD Ohm's law} are reduced to
\begin{subequations}
\begin{align}
  &n_0 h_{\rmi0}\pp{\tilde{\bm{v}}}{t} = -c^2\nbl \tilde{p}_{\rmi} + c\tilde{\bm{J}}\times\bm{B}_0
  \label{e:XMHD e.o.m. linear} \\
  &\lf( \f{h_{\rme0}}{c^2e^2 n_0} \ri)\pp{\tilde{\bm{J}}}{t} = \tilde{\bm{E}} + \f{1}{c}\lf(\tilde{\bm{v}} - \f{\tilde{\bm{J}}}{en_0}\ri)\times\bm{B}_0
  \label{e:XMHD Ohm's law linear}
\end{align}
\end{subequations}
One finds that \eqref{e:XMHD e.o.m. linear} is identical to the momentum equation of the linearized relativistic ideal MHD.
The electron inertia effect appears only in the left hand side of \eqref{e:XMHD Ohm's law linear}.
Note that \eqref{e:XMHD Ohm's law linear} is akin to non-relativistic XMHD~\cite{Kimura2014,KeramidasCharidakos2014,Abdelhamid2015,Lingam2016}, but the left hand side in \eqref{e:XMHD Ohm's law linear} vanishes in the non-relativistic limit because we are focusing on proton inertial scales $k d_\rmi \sim \mu^{-1/4}$.  

Next, we combine \eqref{e:XMHD e.o.m. linear}-\eqref{e:XMHD Ohm's law linear} with Maxwell's equations \eqref{e:Maxwell} and Fourier transform the perturbed fields by $\tilde{f} \propto \exp(\rmi\bm{k}\cdot\bm{x} - \omega t)$.
Below, we assume the equation of state for ideal gas for protons (namely, $h_{\rmi0} = m_\rmi c^2 + [\Gamma/(\Gamma - 1)]T_{\rmi0}$ where $\Gamma = 4/3$ is a specific heat ratio in ultrarelativistic case~\cite{Taub1948}).
Nonetheless, the following results do not change for other choices of the equation of state.
We also omit the entropy wave $\omega = 0$.

Finally, after straightforward, yet cumbersome, manipulations, we obtain the dispersion relation of RXMHD
\begin{strip}
\begin{multline}
  k^2\Bigg\{ \bigg[1 - \check{h}_\rme(k d_\rmi)^2\big((\omega/ck)^2 - 1\big)\bigg]\Big[1 + \sigma - \check{h}_\rme(kd_\rmi)^2(\omega/kc)^2\Big]\omega^2 - \sigma c^2 k_\|^2\Bigg\} \\
  \times\Bigg\{ \Big[1 - \check{h}_\rme(k d_\rmi)^2\big((\omega/ck)^2 - 1\big)\Big]\omega^2\big(\omega^2 - k^2 C_\rmS^2\big) + \sigma\big(\omega^2 - k_\|^2 C_\rmS^2\big)\big(\omega^2 - k^2 c^2\big) \Bigg\} - \lf(1 + \sigma\ri)^2\f{\delta_\rmi^2V_\rmA^2 \omega^2}{c^4}\\
 \times\big(\omega^2 - k^2 c^2\big)\big(\omega^2 - k^2 C_\rmS^2\big) \Bigg\{\Big[1 - \check{h}_\rme(kd_\rmi)^2(\omega/kc)^2\Big]\big(\omega^2 - k^2 c^2\big)k_\|^2 + \Big[1 - \check{h}_\rme(k d_\rmi)^2\big((\omega/ck)^2 - 1\big)\Big]\omega^2k_\+^2\Bigg\} = 0, 
  \label{e:RXMHD dispersion relation}
\end{multline}
\end{strip}
\hskip-0.3em where $\check{h}_\rme = \mu h_\rme/m_\rme c^2$, $\delta_\rmi = h_{\rmi0}/[(4\pi n_0 h_{\rmi0} + B_0^2)e^2 ]^{1/2}$ is the modified proton skin depth~\cite{Kawazura2017b}, $V_\rmA = cB_0/[4\pi(n_0 h_{\rmi0} + B_0^2)]^{1/2}$ is the Alfv\'en speed, and $C_\rmS = c(\Gamma p_{\rmi0}/n_0 h_{\rmi0})^{1/2}$ is the sound speed.
The subscript $\|$ ($\+$) denotes the parallel (perpendicular) component to $\bm{B}_0$.
The right hand side is originated from the Hall effect while the terms multiplied by $\check{h}_\rme$ are coming from the ETI effect.
One finds that $\check{h}_\rme$ is always multiplied by $(kd_\rmi)^2$ in \eqref{e:RXMHD dispersion relation}, and $\check{h}_\rme(kd_\rmi)^2$ is $\sim 1$ because $h_\rme \sim \mu^{-1/2}$ and $kd_\rmi \sim \mu^{-1/4}$.
Ignoring the ETI effect by $\check{h}_\rme \to 0$, one retrieves the relativistic Hall MHD dispersion relation~\cite{Kawazura2017b}
\begin{align}
  &(\omega^2 - k_\para^2 V_\rmA^2) \Bigg\{ \omega^4 - \bigg[ k^2\lf(V_\rmA^2 + \f{1}{1 + \sigma}C_\rmS^2\ri) \nonumber \\
  &\span\hfill+ c^{-2}C_\rmS^2V_\rmA^2k_\para^2 \bigg]\omega^2 + k^2k_\para^2 V_\rmA^2 C_\rmS^2 \Bigg\}  \nonumber\\
  &\span\hfill = \f{\delta_\rmi^2 V_\rmA^2\omega^2}{c^4} \lf( \omega^2 - k_\para^2c^2 \ri)\lf( \omega^2 - k^2c^2 \ri)\lf( \omega^2 - k^2 C_\rmS^2 \ri).
  \label{e:RHMHD dispersion relation}
\end{align}
The non-relativistic limit of \eqref{e:RHMHD dispersion relation} is the dispersion relation of non-relativistic Hall MHD~\cite{Hameiri2005} (which is identical to the non-relativistic limit of \eqref{e:RXMHD dispersion relation} because ETI automatically disappears in the non-relativistic limit).

One finds that \eqref{e:RXMHD dispersion relation} is utterly different from the dispersion relation of non-relativistic XMHD~\cite{Abdelhamid2017b}
\begin{multline}
  \lf(\omega^2 - \f{k_\para^2 V_\rmA^2}{1 + (kd_\rme)^2}\ri)\Bigg[ \omega^4 - k^2\lf(\f{V_\rmA^2}{1 + (kd_\rme)^2} + C_\rmS^2\ri)\omega^2 \\
  + k^2k_\para^2 \f{C_\rmS^2 V_\rmA^2}{1 + (kd_\rme)^2} \Bigg] = \f{d_\rmi^2 k^2k_\para^2 V_\rmA^2 \omega^2}{1 + (kd_\rme)^2}\lf( \omega^2 - k^2C_\rmS^2 \ri). 
  \label{e: NR XMHD dispersion relation}
\end{multline}
In the non-relativistic XMHD, the electron rest mass inertia merely modifies the Alfv\'en speed as $V_\rmA \to V_\rmA/[1 + (kd_\rme)^2]^{1/2}$.
On the other hand, in RXMHD, ETI dramatically changes the wave properties --- which we explain below.

\section{Wave Properties} \label{s:waves}
The most striking difference between RXMHD dispersion relation~\eqref{e:RXMHD dispersion relation} and the non-relativistic XMHD counterpart~\eqref{e: NR XMHD dispersion relation} is that~\eqref{e:RXMHD dispersion relation} is six order with respect to $\omega^2$ while~\eqref{e: NR XMHD dispersion relation} is third order.  
This means that RXMHD has three additional waves in addition to the standard MHD wave family. 
As we found in our previous work~\cite{Kawazura2017b}, one of the three additional waves is originated from the relativistic Hall effect because the right hand side of~\eqref{e:RHMHD dispersion relation} is fourth order with respect to $\omega^2$. 
Here, we call this a Hall wave.
In our previous study, we showed that the Hall wave is superluminous; 
more specifically, as $kd_\rmi$ increases from 0 to infinity, the phase velocity $\bm{v}_\mr{ph} = (\omega/k^2)\bm{k}$ decreases from infinity to $c$, and the group velocity $\bm{v}_\mr{gr} = \p\omega/\p\bm{k}$ increases from 0 to $c$.
This means that the Hall wave is indistinguishable from light at small scales as long as one views the propagation properties.

\begin{figure}
  \begin{center}
    \includegraphics*[width=0.47\textwidth]{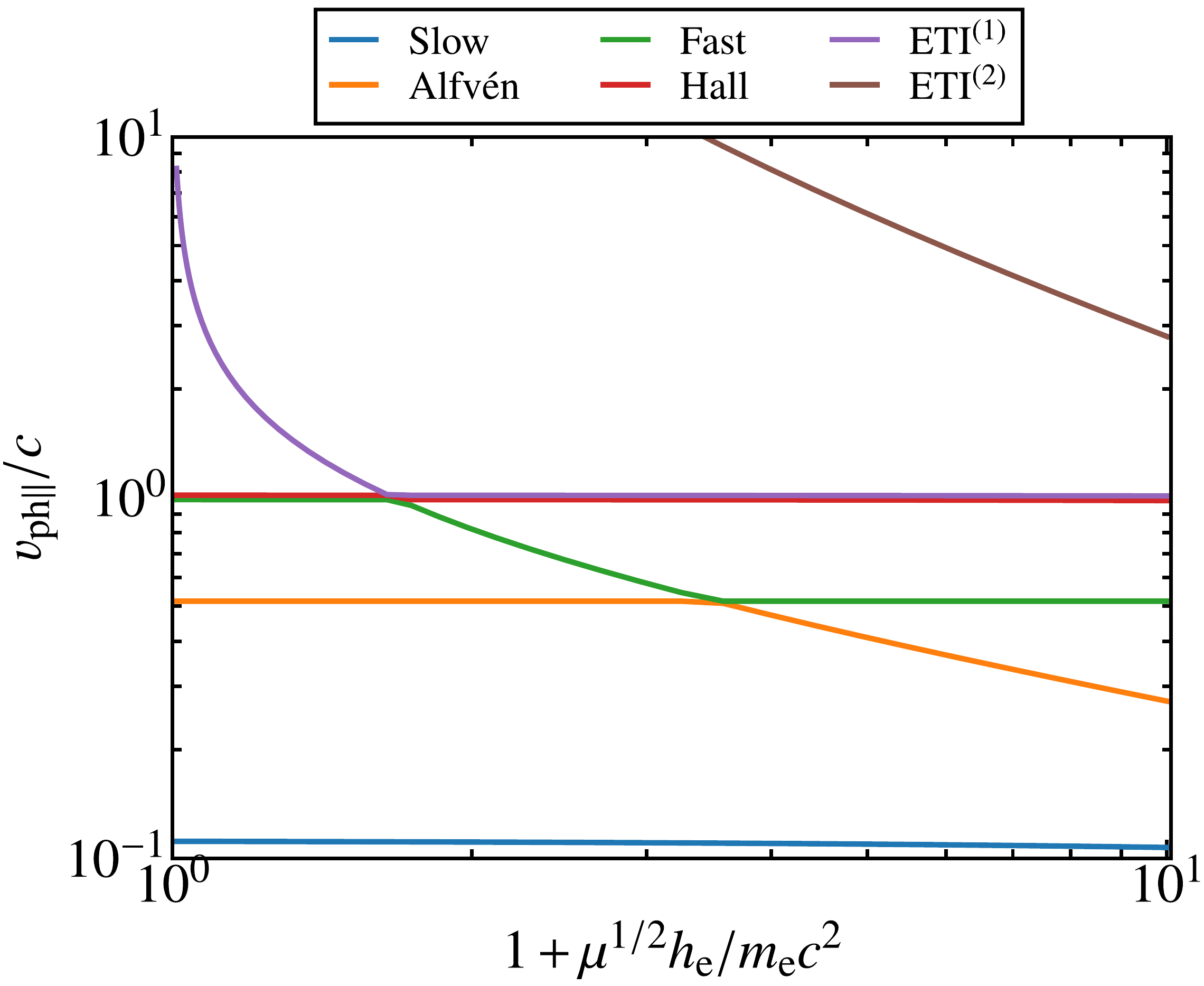}
  \end{center}
  \caption{Dependence of parallel phase velocity on $h_\rme$ when $kd_\rmi = 8$, $\sigma = 4$, and $T_{\rmi0}/m_\rmi c^2 = 1$.}
  \label{f:vpz vs he}
\end{figure}

When ETI is finite, we have two more waves, which we call ETI$^{(1)}$ and ETI$^{(2)}$ waves in the order of phase speed.
These new waves are generated because ETI couples with the displacement current, as is evident from the fact that $h_\rme$ is always paired with $\omega/ck$ in~\eqref{e:RXMHD dispersion relation}.
To see the order relation of phase speed of six RXMHD waves and their asymptotic behavior against $h_\rme$, we plot the parallel component of phase velocity of all wave solutions vs. $h_\rme$ for $kd_\rmi = 8$, $\sigma = 4$, and $T_{\rmi0}/m_\rmi c^2 = 1$ in Fig.~\ref{f:vpz vs he}. 
We find that the phase speed of two ETI waves are greater than the Hall wave.
Both the Hall wave and two ETI waves are superluminous with the phase velocity decreasing from infinity to $c$ as $h_\rme$ increases from 0 to infinity.
We also find that the phase velocity of fast magnetosonic and Alfv\'en waves decreases as $h_\rme$ increases while that of the slow magnetosonic wave is almost constant.

\begin{figure*}
  \begin{center}
    \includegraphics*[width=1.0\textwidth]{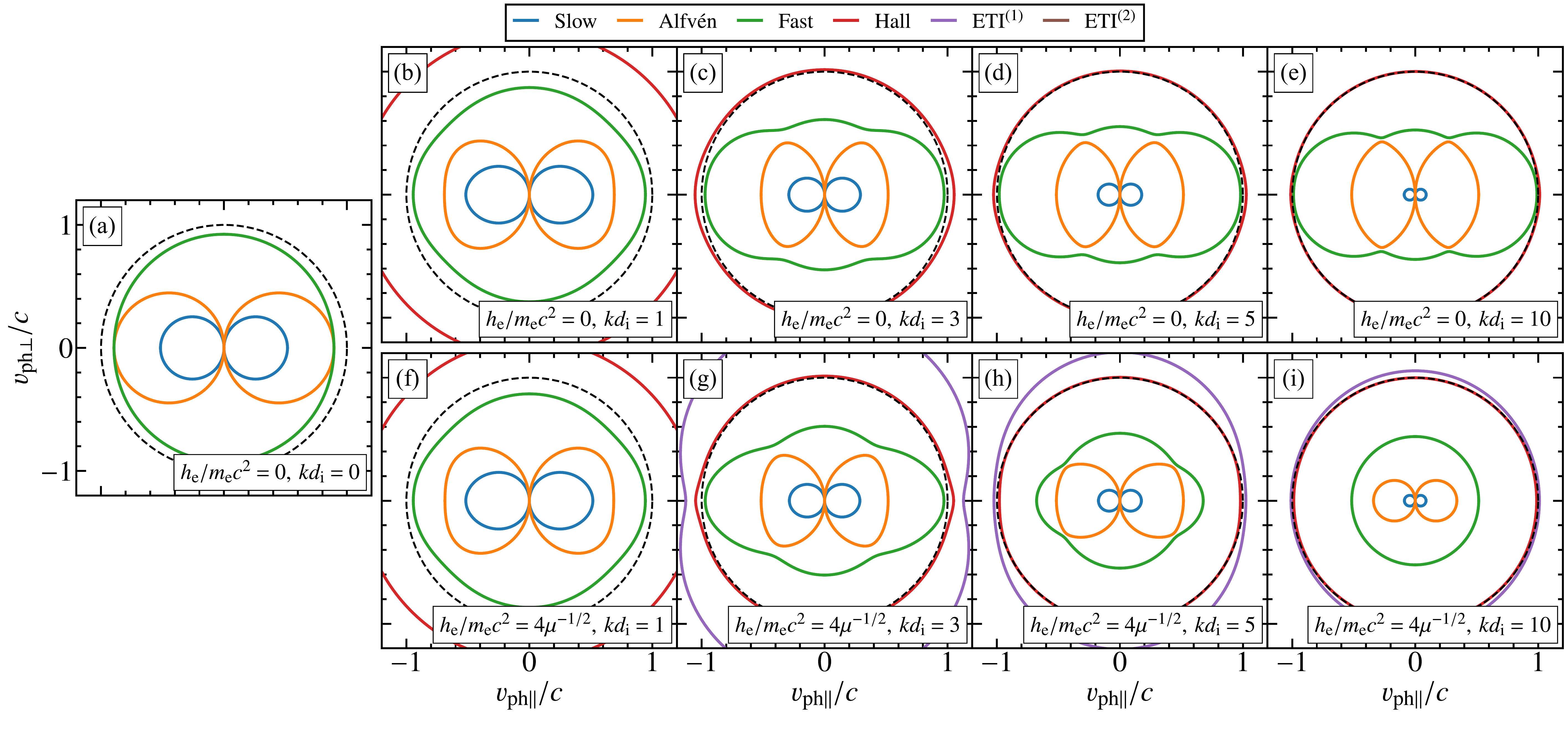}
  \end{center}
  \caption{The phase diagram for (a) relativistic ideal MHD with $kd_\rmi = 0$, (b-e) RHMHD with $h_\rme = 0$, and (f-i) RXMHD with $h_\rme/m_\rme c^2 = 4\mu^{-1/2}$, where $\sigma$ and $T_{\rmi0}/m_\rmi /c^2$ are fixed at 4 and 1, and $kd_\rmi$ varies as $kd_\rmi = (0,\, 1,\, 3,\, 5, \, 10)$ from left to right panels. The broken circles indicate the speed of light.}
  \label{f:phase_diagram_sgm4}
\end{figure*}

To explore the wave properties in all directions, we draw the phase diagram (a trajectory of $\bm{v}_\mr{ph}$) in Figure~\ref{f:phase_diagram_sgm4} with various $kd_\rmi = (0,\, 1,\, 3,\, 5, \, 10)$ for RHMHD ($h_\rme =0 $) and RXMHD ($h_\rme/m_\rme c^2 = 4\mu^{-1/2}$), where $\sigma$ and $T_{\rmi0}/m_\rmi /c^2$ are fixed at 4 and 1, respectively (see also Supplementary material~\ref{f:phase_diagram_sgm4_full} for a finer interval of $kd_\rmi$).
First, we summarize the features of HMHD waves that were presented in our previous study~\cite{Kawazura2017b}; the phase velocity of Hall wave is almost perfectly isotropic. 
Moreover, the phase diagram of the fast wave becomes nearly oval shape with $v_\mr{ph||} \simeq 2v_\mr{ph\+}$ as $kd_\rmi$ increases, while that in non-relativistic HMHD becomes dumbbell shape with $v_\mr{ph||} \gg 2v_\mr{ph\+}$~\cite{Hameiri2005}, meaning that the relativistic Hall effect isotropize the fast wave. 
In the case of RXMHD, the phase diagram of the fast wave is even more isotropic at large $kd_\rmi$ [Fig.~\ref{f:phase_diagram_sgm4} (i)], and the shapes of fast and Alfv\'en waves at large $kd_\rmi$ are akin to those of relativistic ideal MHD ($kd_\rmi = 0$).
One finds that the phase diagram of ETI$^{(1)}$ wave is slightly anisotropic with $v_\mr{ph\+} > v_\mr{ph\|}$ at $kd_\rmi \lesssim \mu^{-1/4}$ [Fig.~\ref{f:phase_diagram_sgm4} (g)], but it gets isotopic at large $kd_\rmi$ because ETI$^{(1)}$ wave converges to light.

\begin{figure*}
  \begin{center}
    \includegraphics*[width=1.0\textwidth]{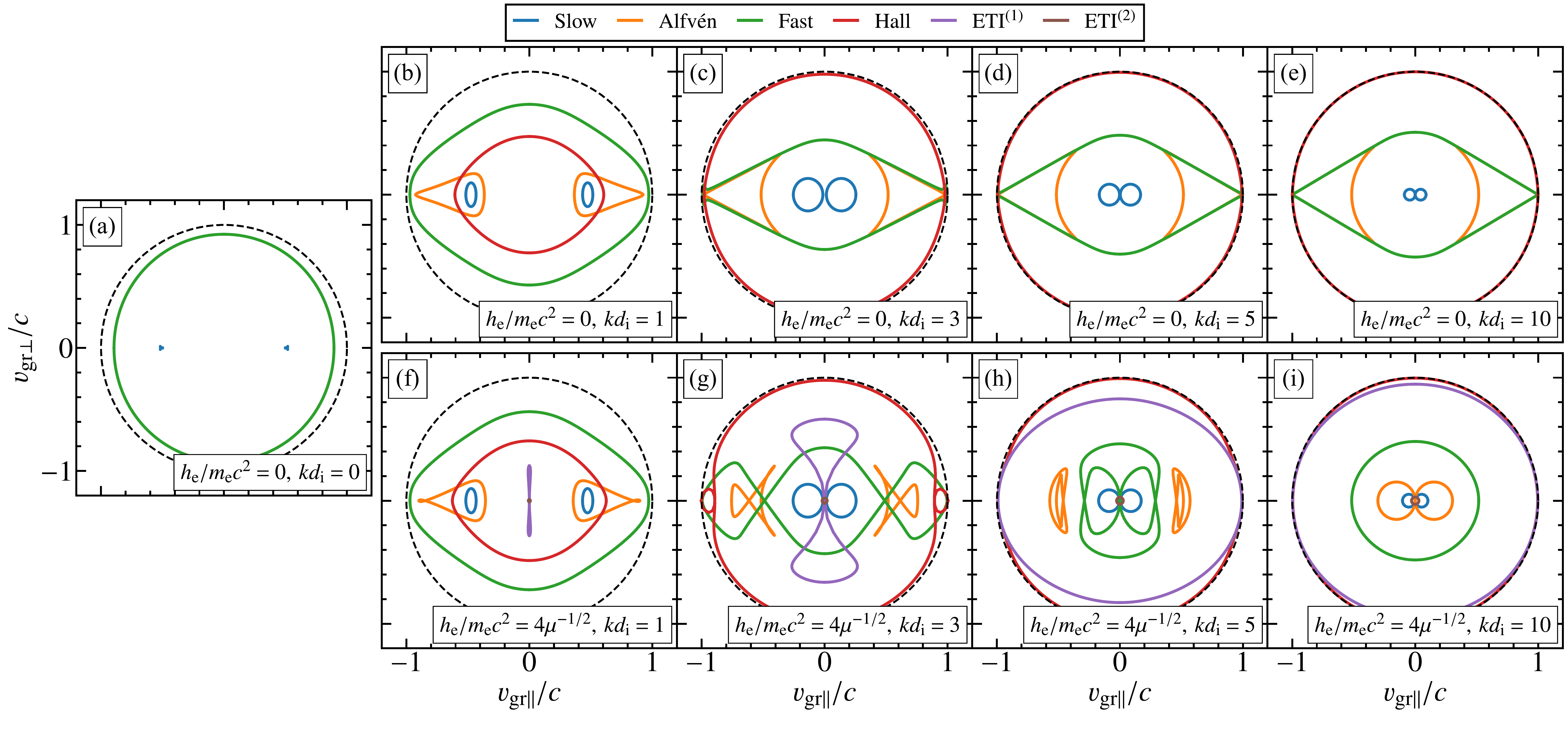}
  \end{center}
  \caption{The group diagram for (a) relativistic ideal MHD with $kd_\rmi = 0$, (b-e) RHMHD, and (f-i) RXMHD. The setting is the same as that for Fig.~\ref{f:phase_diagram_sgm4}.}
  \label{f:group_diagram_sgm4}
\end{figure*}

Next, we show the group diagram (a trajectory of $\bm{v}_\mr{gr}$, also known as ray surfaces) with the same parameter setting as Fig.~\ref{f:phase_diagram_sgm4}~(see also Supplementary material~\ref{f:group_diagram_sgm4_full} for a finer interval of $kd_\rmi$).
Drawing a group diagram is important as it graphically represents the shape of the wave front and the direction of energy flow~\cite[see][for more details]{Goedbloed2004,Hameiri2005}.
As we found in our previous work~\cite{Kawazura2017b}, the way relativistic Hall effect modifies the group diagram is vastly different from the way non-relativistic Hall effect does~\cite{Hameiri2005}. 
In non-relativistic HMHD, the group velocity of the fast wave is much greater than that of Alfv\'en wave in all direction at large $kd_\rmi$.
On the other hand, in RHMHD, the group velocity surface of the fast wave mostly overlaps with that of Alfv\'en wave (which we call ``coalescence'') [Fig.~\ref{f:group_diagram_sgm4} (c-e)].
The coalescence happens because $\bm{v}_\mr{gr}$ of fast waves is limited by the speed of light, and thus it cannot be deviated from the group surface of Alfv\'en wave.
Now, when ETI is finite, we find that the group diagram is significantly changed.
First, comparing Fig.~\ref{f:group_diagram_sgm4} (c) and (g), one finds that the group velocity surfaces of fast and Alfv\'en waves become much more complicated when ETI is finite.
Second, the ETI$^{(1)}$ wave propagates anisotropically in a certain wavenumber range, and it primarily propagates in the perpendicular direction to $\bm{B}_0$ [Fig.~\ref{f:group_diagram_sgm4} (g)]. 
Third, the group velocity of Alfv\'en wave is greater than the fast wave when $kd_\rmi$ is not too large [Fig.~\ref{f:group_diagram_sgm4} (h)].
Lastly, when $kd_\rmi$ is large, the coalescence of fast and Alfv\'en wave is broken; more specifically, the group velocity of the fast wave is greater than that of Alfv\'en wave in all directions when [Fig.~\ref{f:group_diagram_sgm4} (i)].
Comparing Fig.~\ref{f:phase_diagram_sgm4} (i) and Fig.~\ref{f:group_diagram_sgm4} (i), one also finds that the phase and group diagrams of RXMHD are almost identical when $kd_\rmi$ is large. 
Note that the group velocity of the ETI$^{(2)}$ wave is almost zero, and thus it does not play any meaningful role in this parameter range (yet, it can be meaningful in even smaller scales, but the electron rest mass inertia would further change the wave properties).

\begin{figure*}
  \begin{center}
    \includegraphics*[width=1.0\textwidth]{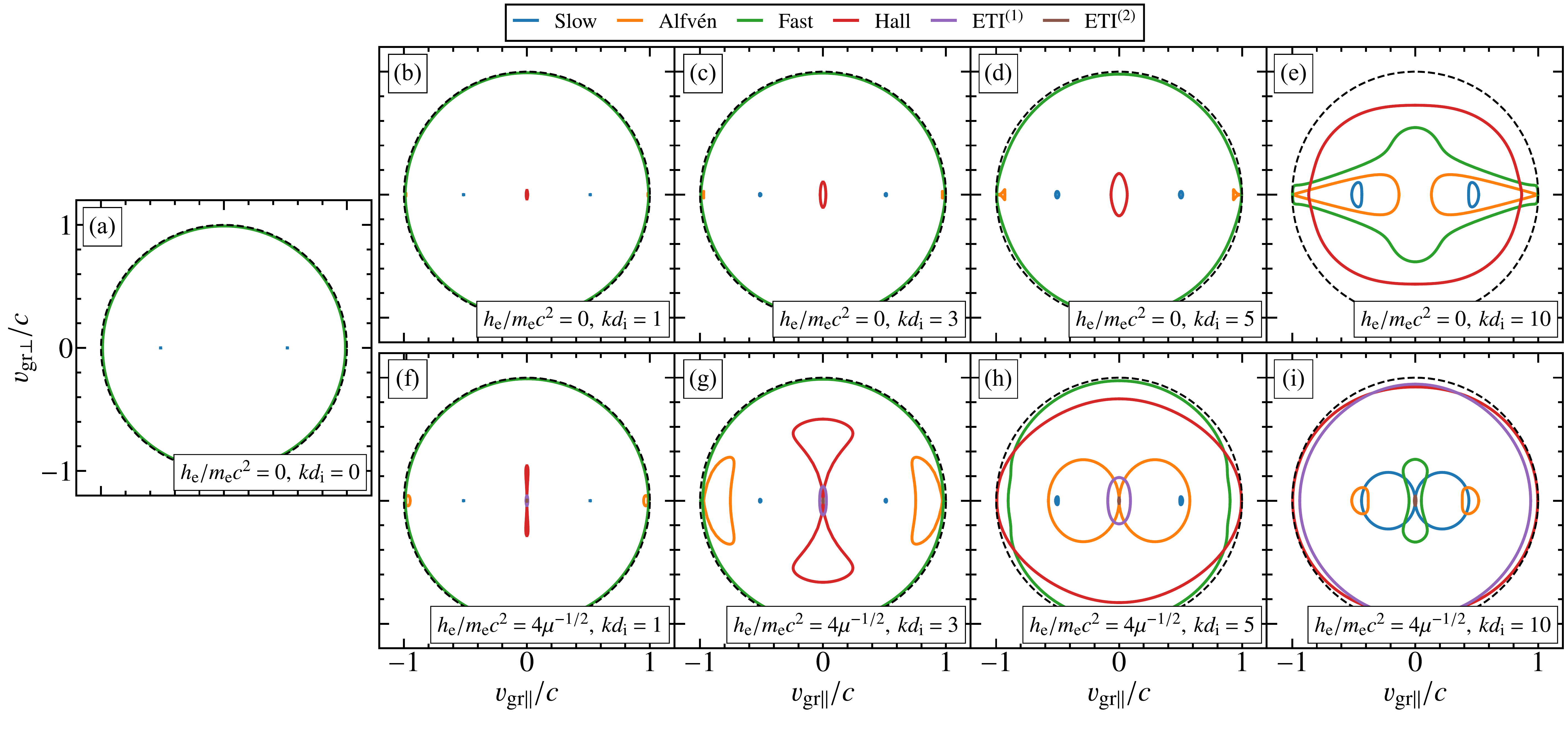}
  \end{center}
  \caption{The same as Fig.~\ref{f:group_diagram_sgm4}, but for $\sigma = 40$.}
  \label{f:group_diagram_sgm40}
\end{figure*}

Finally, we explore even strongly magnetized case.
In RHMHD, as one can see from \eqref{e:RHMHD dispersion relation}, the scale at which the relativistic Hall effects switches on is not $d_\rmi$ but $\delta_\rmi$~\cite{Kawazura2017b}.
Since $\delta_\rmi$ gets smaller as the magnetic energy gets larger, the wave properties in RHMHD are almost the same as those of relativistic ideal MHD.
However, this behavior does not apply to RXMHD because the terms containing ETI in \eqref{e:RHMHD dispersion relation} do not depend on $\sigma$.
This is evident in Fig.~\ref{f:group_diagram_sgm40} which shows the group velocity diagram for $\sigma = 40$ case. 
The diagrams of RHMHD [Fig.~\ref{f:group_diagram_sgm40} (b)-(d)] are almost the same as relativistic ideal MHD [Fig.~\ref{f:group_diagram_sgm40} (a)] unless $kd_\rmi$ is sufficiently large.
On the other hand, in the case of RXMHD, the group diagram is significantly different from that of relativistic ideal MHD.  
Figure~\ref{f:group_diagram_sgm40} (g) shows that the Hall wave propagates anisotropically in the same manner as the ETI$^{(1)}$ wave with $\sigma = 4$ [Fig.~\ref{f:group_diagram_sgm4} (g)].
Even more interestingly, Fig.~\ref{f:group_diagram_sgm40} (i) shows that the fast wave group velocity can be smaller than that of slow wave for the $\sigma = 40$ case while the slow wave is always smallest in the case of $\sigma = 4$ [Fig.~\ref{f:group_diagram_sgm4} (i)].


\section{Concluding remarks} \label{s:conclusion}
In this paper, we have derived the linear dispersion relation of RXMHD when the electrons are ultrarelativistic (i.e., $T_\rme/m_\rme c^2 \sim \mu^{-1/2}$) and wavelength is slightly shorter than ion skin depth (i.e., $kd_\rmi \sim \mu^{-1/4}$).
We have shown that ETI significantly modifies the properties of waves. 
One wave emerges due to the Hall effect, and two waves emerge due to ETI, all of which are superluminous.
These waves can propagate anisotropically in a perpendicular direction to $\bm{B}_0$ at certain value of $kd_\rmi$.
Moreover, the phase and group diagrams of three standard MHD waves are distorted by ETI;
the fast wave is isotropized, and the coalescence of fast and Alfv\'en waves are broken at large $kd_\rmi$. 
Finally, there is a range of wavenumber where the group velocity of fast wave is smaller than that of the Alfv\'en wave for moderate value of $\sigma$ and smaller than that of slow wave for high $\sigma$.

\section{Acknowledgements}
The author is indebted to Shigeo Kimura and Kenji Toma for useful suggestions.
This work was supported by JSPS KAKENHI grant JP20K14509. 


\bibliographystyle{elsarticle-num}

\bibliography{references}

\clearpage

\setcounter{figure}{0}
\renewcommand{\figurename}{Supplementary material}
\begin{figure*}[b]
  \begin{center}
    \includegraphics*[width=0.85\textwidth]{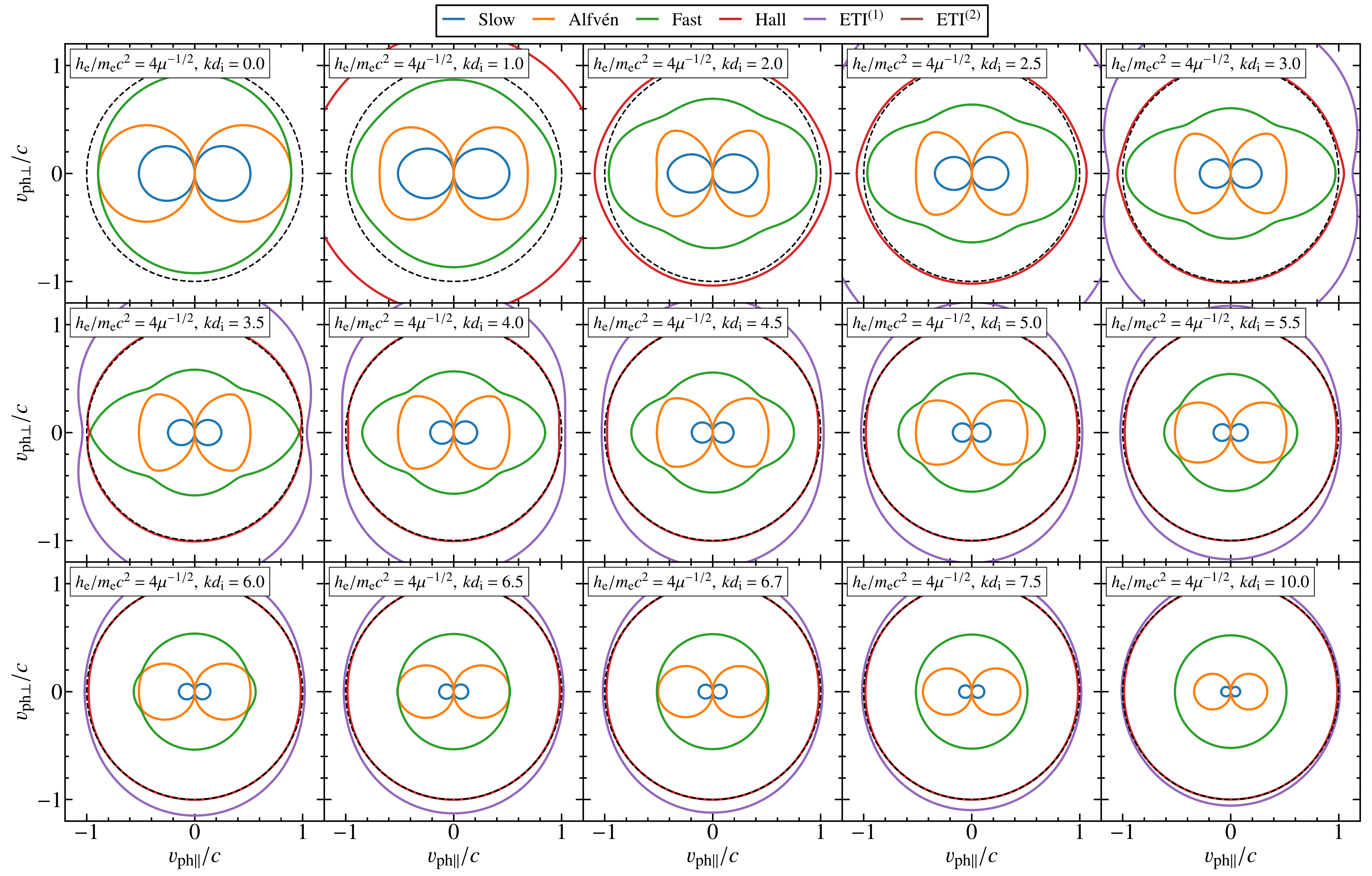}
  \end{center}
  \caption{The phase diagram of RXMHD with $\sigma = 4$ and a finer interval of $kd_\rmi$.}
  \label{f:phase_diagram_sgm4_full}
\end{figure*}
\begin{figure*}[b]
  \begin{center}
    \includegraphics*[width=0.85\textwidth]{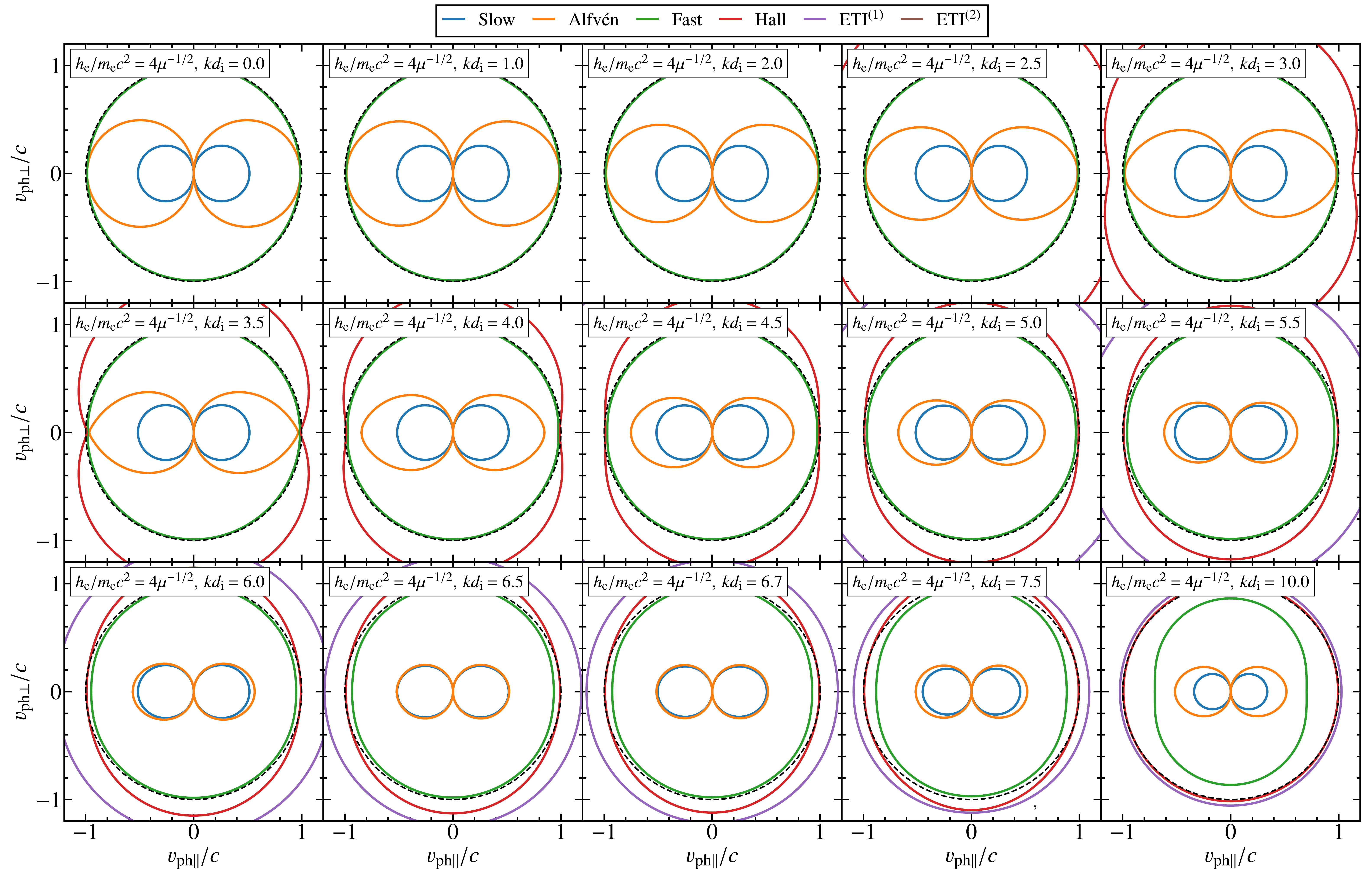}
  \end{center}
  \caption{The phase diagram of RXMHD with $\sigma = 40$ and a finer interval of $kd_\rmi$.}
  \label{f:phase_diagram_sgm40_full}
\end{figure*}
\begin{figure*}[b]
  \begin{center}
    \includegraphics*[width=0.85\textwidth]{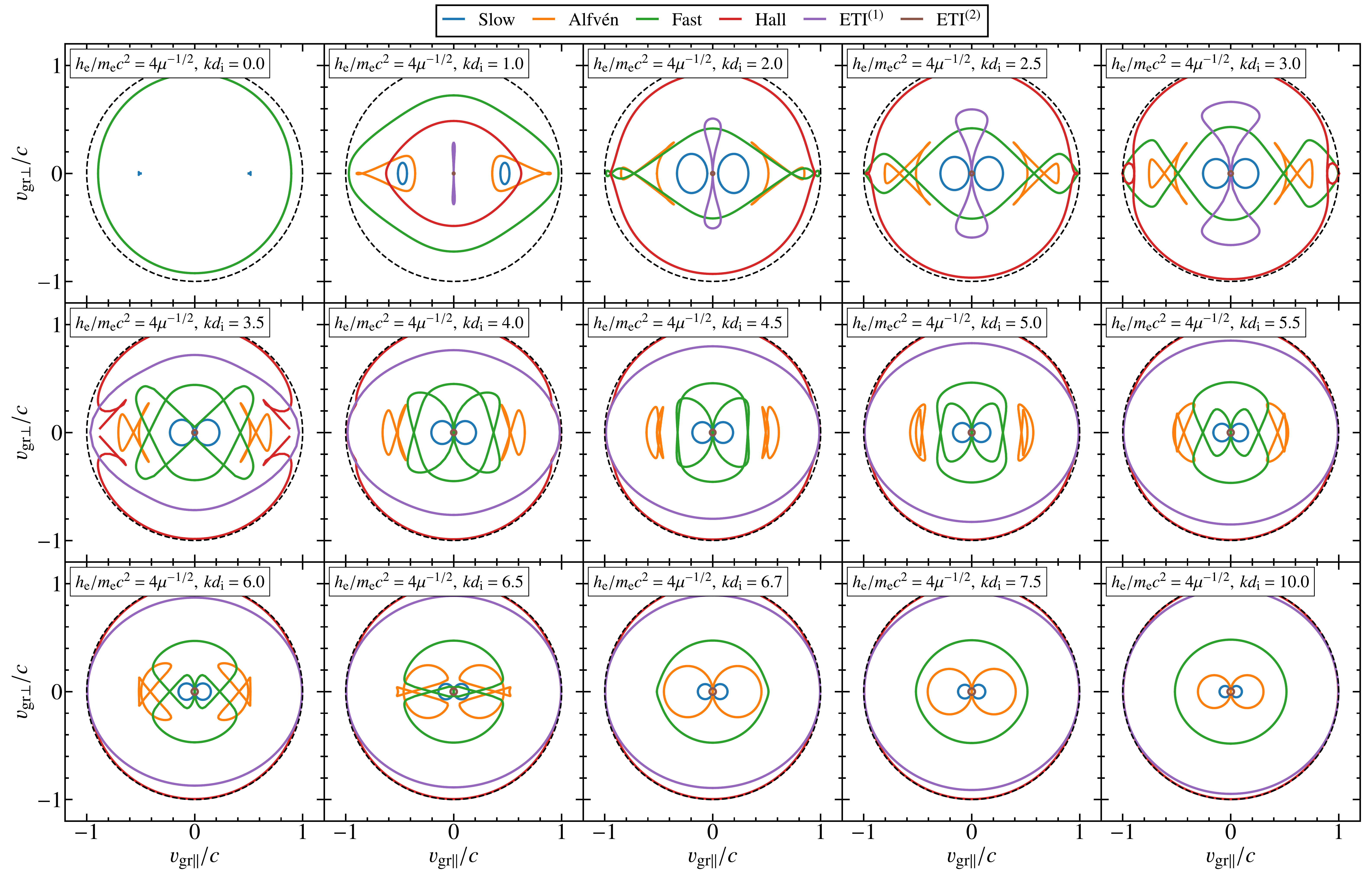}
  \end{center}
  \caption{The group diagram of RXMHD with $\sigma = 4$ and a finer interval of $kd_\rmi$.}
  \label{f:group_diagram_sgm4_full}
\end{figure*}
\begin{figure*}[b]
  \begin{center}
    \includegraphics*[width=0.85\textwidth]{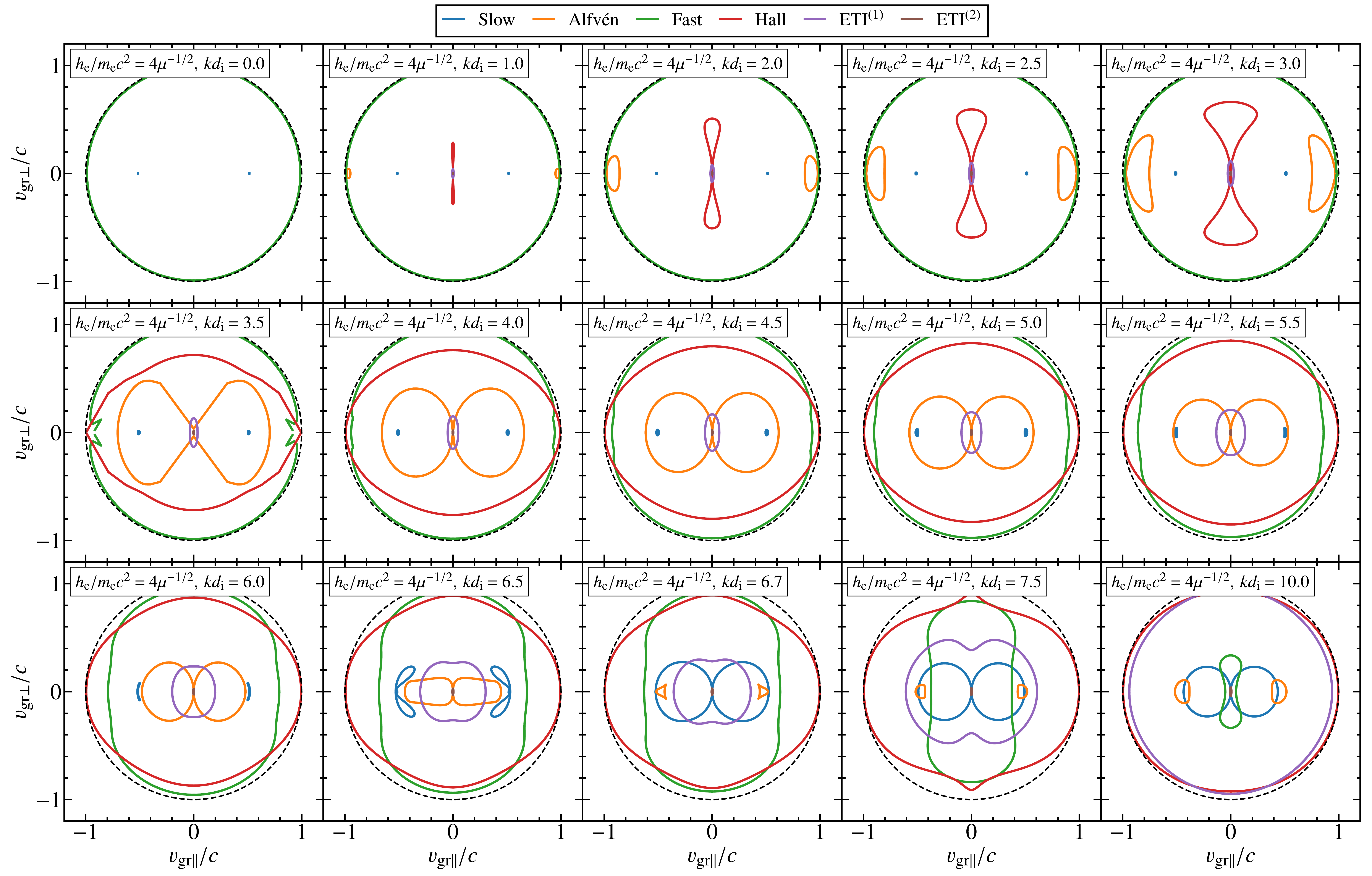}
  \end{center}
  \caption{The group diagram of RXMHD with $\sigma = 40$ and a finer interval of $kd_\rmi$.}
  \label{f:group_diagram_sgm40_full}
\end{figure*}

\end{document}

%% file: preamble.tex
\usepackage{amssymb,amsmath,framed}
\usepackage{graphicx}
\usepackage{epstopdf, epsfig}
\usepackage[dvipsnames]{xcolor}
\usepackage{cuted}

\usepackage[numbers,sort&compress]{natbib}

\usepackage[scaled=0.9]{helvet} 
\usepackage{stix2}
\newcommand{\bm}[1]{{\boldsymbol {\mathrm #1}}} 


\newcommand{\p}{\partial}
\newcommand{\f}[2]{\frac{#1}{#2}}
\newcommand{\mr}[1]{\mathrm{#1}}
\newcommand{\lf}{\left}
\newcommand{\ri}{\right}

\newcommand{\pp}[2]{\frac{\p #1}{\p #2}}

\newcommand{\nbl}{\nabla}
\newcommand{\para}{{||}}
\newcommand{\+}{{\perp}}

\makeatletter
\newcommand{\vast}{\bBigg@{4}}
\newcommand{\Vast}{\bBigg@{5}}

\makeatother

\DeclareMathAlphabet{\mathpzc}{OT1}{pzc}{m}{it}

\newcommand{\rme}{\mathrm{e}}

\newcommand{\rmi}{\mathrm{i}}

\newcommand{\rmA}{\mathrm{A}}

\newcommand{\rmS}{\mathrm{S}}

